\documentclass[10pt]{scrartcl}

\usepackage[utf8]{inputenc}
\usepackage[margin=12mm]{geometry}
\usepackage[svgnames]{xcolor}

\usepackage{enumitem,multicol}
\setlist{leftmargin=*}
\usepackage[format=plain,bf,sf,hypcap=false]{caption}

\usepackage{mathtools,amssymb,tensor,siunitx,dsfont}

\usepackage[colorlinks=true,allcolors=DarkBlue]{hyperref}
\usepackage{cleveref}

\usepackage[backend=biber,sorting=none,citestyle=numeric-comp,giveninits=true,doi=false]{biblatex}
\renewbibmacro{in:}{}
\AtEveryBibitem{\clearfield{number}}

\DeclareFieldFormat[article]{volume}{\mkbibbold{#1}} 
\DeclareFieldFormat{pages}{#1}

\DeclareFieldFormat{eprint:arxiv}{\href{http://arxiv.org/abs/#1}{#1}}

\newcommand{\dif}{\mathrm{d}}
\renewcommand{\vec}[1]{\boldsymbol{#1}}
\DeclareMathOperator{\tr}{tr}
\DeclareMathOperator{\Tr}{Tr}

\setkomafont{section}{\large}

\begin{document}

\title{Graviton fluctuations erase the cosmological constant}
\author{C. Wetterich\footnote{\href{mailto:c.wetterich@thphys.uni-heidelberg.de}{c.wetterich@thphys.uni-heidelberg.de}}}
\date{\small Universität Heidelberg, Institut für Theoretische Physik, Philosophenweg 16, D-69120 Heidelberg}
\maketitle

\begin{abstract}
    \noindent\small Graviton fluctuations induce strong non-perturbative infrared renormalization effects for the cosmological constant. The functional renormalization flow drives a positive cosmological constant towards zero, solving the cosmological constant problem without the need to tune parameters. We propose a simple computation of the graviton contribution to the flow of the effective potential for scalar fields. Within variable gravity we find that the potential increases asymptotically at most quadratically with the scalar field. With effective Planck mass proportional to the scalar field, the solutions of the derived cosmological equations lead to an asymptotically vanishing cosmological ``constant'' in the infinite future, providing for dynamical dark energy in the present cosmological epoch. Beyond a solution of the cosmological constant problem, our simplified computation also entails a sizeable positive graviton-induced anomalous dimension for the quartic Higgs coupling in the ultraviolet regime, substantiating the successful prediction of the Higgs boson mass within the asymptotic safety scenario for quantum gravity.
\end{abstract}

\begin{multicols}{2}

The tiny value of the cosmological constant $V$, as compared to the Fermi scale of weak interactions, $h_0 = \SI{176}{\giga\electronvolt}$, or the Planck mass $M = \SI{2.4e18}{\giga\electronvolt}$, is an old puzzle \cite{weinberg1989cosmological}. In units of the Planck mass this value amounts to $V/M^4 \approx \num{e-120}$. Within quantum field theory the cosmological constant can be identified with the value that the effective potential of scalar fields takes at its minimum or for a given cosmological solution. Approximating the effective potential for the Higgs doublet $h$ by $U = \lambda_h \, (h^\dagger h)^2/2 - \lambda_h \, h_0^2 \, h^\dagger h + c_h$, it is hard to understand why the difference $V = c_h - \lambda_h h_0^4/2$ should take a value $\sim (\SI{2e-3}{\electronvolt})^4$, much smaller than $h_0^4$. In this note, we argue that quantum gravity effects  induce a strong renormalization of the cosmological constant, making its value on cosmological scales almost independent of the value of $c_h$ at the Fermi scale.

In a general quantum field theory the cosmological constant or the effective scalar potential are functions of an infrared (IR) scale $k$. For a given $k$ only fluctuations with momenta larger than $k$, or wavelength smaller than $k^{-1}$, are included in the computation of the renormalized quantities. For example, $k$ may be set by the momentum of particles in a scattering process. For this case, the $k$-dependence or ``running'' of the strong coupling constant in quantum chromodynamics has been impressively confirmed by observation. We find that graviton fluctuations are responsible for a strong running of $V(k)$. Even if $V(k)$ is of the order $h_0^4$ or larger at the scale $k = h_0$, it will take a tiny value on a cosmological scale $k \approx H$, as required by observation.

A widespread view asserts that quantum gravity effects become important on short length scales of the order of the Planck length $l_\text{P} = 1/M$, while being negligible for much larger wavelengths of the fluctuations. One would then expect that quantum gravity effects can give only negligible contributions to the running of $V$ between $k = h_0$ and $k = H$, suppressed by a factor $M^{-2}$. We will see that these simple considerations do not hold due to a potential instability of the graviton fluctuations. While the contribution of graviton fluctuations indeed involves a factor $M^{-2}$, there is also an enormous enhancement factor for the flow of $V$ in the vicinity of the instability. As the instability is approached, the flow equations become singular, preventing in this way any unstable behavior. This ``avoidance of instabilities'' gives rise to strong IR effects in quantum gravity.

Our method for the computation of the quantum gravity effects for the cosmological constant is based on the exact flow equation for the effective average action \cite{wetterich1993exact}. Applied to the ultraviolet (UV) behavior of quantum gravity \cite{reuter1998nonperturbative}, it has already led to substantial evidence for asymptotic safety \cite{weinberg1979ultraviolet}, which would render quantum gravity non-perturbatively renormalizable. In this note, we concentrate on the IR effects of quantum gravity. While rather extensive information has been available on the UV flow for a long time \cite{dou1998running,souma1999non,reuter2002renormalization,litim2004fixed,bonanno2017asymptotically,manrique2011asymptotically}, a detailed exploration of the IR flow is only at its beginning \cite{christiansen2014fixed,christiansen2016global,henz2013dilaton,henz2017scaling,christiansen2015local,denz2016towards}.

Due to the importance of the gauge symmetry of diffeomorphism transformations, a full computation of the flow equation for $V$ or $U$ gets rather involved already for simple truncations of the effective average action. Our aim is here to demonstrate the key features of the gravitationally induced flow of the effective potential. For this purpose, we only include the contributions from the graviton fluctuations, leaving other degrees of freedom in the metric and ghosts aside. This leads to an intuitive and simple expression for the flow equation. The restriction to the graviton fluctuations also avoids several conceptual and technical complications. The graviton fluctuations are physical fluctuations that do not depend on the choice of a gauge. Analytic continuation between Euclidean and Minkowski signature is particularly simple for the graviton fluctuations. Furthermore, the graviton fluctuations cannot mix with scalar or vector fluctuations for any geometry with rotation invariance. This block diagonal form of the propagator matrix allows us to treat the graviton fluctuation effects independently of other fluctuations.

Our proposal for the solution of the cosmological constant problem will be formulated within variable gravity, with effective Planck mass $M(\chi)$ or Newton's ``constant'' depending on a scalar field $\chi$. The main ingredient states that the strong IR graviton fluctuations exclude for $\chi \to \infty$ an increase of the effective potential $U(\chi)$ faster than $M^2(\chi)$. The observable cosmological constant is determined by the dimensionless ratio $U(\chi)/M^4(\chi)$, which decays asymptotically $\sim M^{-2}(\chi)$. The cosmological solutions of models with variable Planck mass lead to an asymptotic increase $M(\chi) \to \infty$ for the infinite future, resulting in an asymptotically vanishing cosmological constant. The universal asymptotic properties of $U(\chi)$ are independent of microscopic parameters, as characteristic for an IR fixed point.

The most prominent features of the graviton-induced renormalization effects can already be seen without the scalar field, and we therefore start with simple Einstein gravity in Euclidean flat space. The inclusion of scalar fields and extensions to Minkowski space and other geometries are discussed subsequently. We believe that our simple approach already contains all important elements for the strong IR-gravity effect.

\section{Scale-dependent cosmological constant}

Consider a flat Euclidean background geometry, $\bar g_{\mu\nu} = \delta_{\mu\nu}$. We restrict the metric fluctuations to the transversal traceless part,
\begin{equation}
    \begin{alignedat}{2}
        &\tensor{g}{_\mu_\nu}
        = \tensor{\delta}{_\mu_\nu} + \tensor{t}{_\mu_\nu},
        \qquad
        &&\tensor{t}{_\mu_\nu}
        = \tensor{t}{_\nu_\mu},\\
        &\tensor{t}{^\mu_\mu}
        = \delta^{\mu\nu} t_{\mu\nu}
        = 0,
        \qquad
        &&q^\mu t_{\mu\nu}
        = 0.
    \end{alignedat}
\end{equation}
In momentum space $\tensor{t}{_\mu_\nu}$ is a function of the Euclidean four-momentum $q_\rho = (q_0,\vec{q})$ with $q^2 = q^\rho q_\rho$. The Euclidean Einstein-Hilbert action with cosmological constant $V$, reduced Planck mass $M$ and curvature scalar $\tilde{R}$ reads
\begin{equation}\label{eqn:einstein hilbert action}
    \Gamma
    = \int_x\sqrt{g}\biggl(V - \frac{M^2}{2}\tilde{R}\biggr).
\end{equation}
The term quadratic in $\tensor{t}{_\mu_\nu}$,
\begin{equation}
    \Gamma_2
    = \frac{1}{2} \int_{q,q^\prime} \tensor{t}{_\mu_\nu}(-q^\prime) \, \tensor{\Gamma}{^{(2)}^\mu^\nu^\rho^\tau}(q^\prime,q) \, \tensor{t}{_\rho_\tau}(q),
\end{equation}
defines the second functional derivative
\begin{equation}\label{eqn:inverse propagator}
    \tensor{\Gamma}{^{(2)}^\mu^\nu^\rho^\tau}(q^\prime,q)
    = \biggl(\frac{M^2q^2}{4} - \frac{V}{2}\biggr) \tensor{P}{^{(t)}^\mu^\nu^\rho^\tau}(q) \delta(q^\prime - q).
\end{equation}
Here $\int_q = \int \frac{\dif^4 q}{(2\pi)^4}$, $\delta(q^\prime - q) = (2\pi)^4 \delta^4(q_\mu^\prime - q_\mu)$, and the projector on $\tensor{t}{_\mu_\nu}$, $P^{(t)} = (P^{(t)})^2$ is given by
\begin{equation}\label{eqn:projector}
    \begin{aligned}
        &\tensor*{P}{^{(t)}_\mu_\nu_\rho_\tau}
        = \frac{1}{2} \bigl(\tensor{\tilde{P}}{_\mu_\rho} \tensor{\tilde{P}}{_\nu_\tau} + \tensor{\tilde{P}}{_\mu_\tau} \tensor{\tilde{P}}{_\nu_\rho}\bigr) - \frac{1}{3} \tensor{\tilde{P}}{_\mu_\nu} \tensor{\tilde{P}}{_\rho_\tau},\\
        &\tensor{\tilde{P}}{_\mu^\nu} = \tensor{\delta}{_\mu^\nu} - \frac{q_\mu q^\nu}{q^2},
        \qquad
        \Tr P^{(t)}
        = \tensor*{P}{^{(t)}_\mu_\nu^\mu^\nu} = 5.
    \end{aligned}
\end{equation}

The propagator or connected two-point function for $t_{\mu\nu}$ is the inverse of $\Gamma^{(2)}$ on the projected space \cite{wetterich2016quantum}
\begin{equation}\label{eqn:propagator}
    \tensor{G}{_\mu_\nu_\rho_\tau}(q^\prime,q)
    = \frac{4}{M^2}\biggl(q^2 - \frac{2V}{M^2}\biggr)^{-1} \tensor*{P}{^{(t)}_\mu_\nu_\rho_\tau}(q) \, \delta(q^\prime - q).
\end{equation}
We observe that the cosmological constant $V$ acts like a mass term for the graviton,
\begin{equation}\label{eqn:graviton mass}
    m^2 = -\frac{2V}{M^2}.
\end{equation}
For $V>0$ the propagator is tachyonic and one expects strong infrared instabilities. The negative sign in \cref{eqn:graviton mass} will be the crucial ingredient for the solution of the cosmological constant problem. We recall that for $V \neq 0$ flat space is \textit{not} a solution of the field equation
\begin{equation}
    M^2 \bigl(\tensor{\tilde{R}}{_\mu_\nu} - \tfrac{1}{2} \tilde{R} \, \tensor{g}{_\mu_\nu}\bigr)
    = -V g_{\mu\nu}.
\end{equation}
The metric propagator in a flat background is therefore an ``off-shell propagator'', as needed for the functional renormalization flow.

Functional renormalization investigates the dependence of the effective average action $\Gamma_k$ on an infrared cutoff scale $k$. We work in the simple truncation \labelcref{eqn:einstein hilbert action}, where $V$ and $M^2$ are now $k$-dependent ``running couplings''. We will see that the flow of $V$ with $t = \ln(k/M)$ is such that the potential instability in the propagator \labelcref{eqn:propagator} is avoided. Our starting point is the exact functional flow equation for the effective average action \cite{wetterich1993exact}
\begin{equation}\label{eqn:wetterich}
    \partial_t \Gamma_k
    = \frac{1}{2} \tr \{G_k \, \partial_t R_k\},
\end{equation}
with
\begin{equation}\label{eqn:inverse}
    (\Gamma_k^{(2)} + R_k) \, G_k = \mathds{1}.
\end{equation}
It involves the matrix $\Gamma_k^{(2)}$ of second functional derivatives of $\Gamma_k$ and the IR cutoff that we take as
\begin{equation}\label{eqn:ir cutoff}
    \tensor{R}{_k^\mu^\nu^\rho^\tau}(q^\prime,q)
    = \frac{M^2}{4} \, R_k(q) \, \tensor{P}{^{(t)}^\mu^\nu^\rho^\tau} \, \delta(q^\prime - q).
\end{equation}
The cutoff function will be chosen such that an addition of $R_k(q)$ to to the kinetic term $q^2$ in $\Gamma^{(2)}$ provides a type of positive (momentum-dependent) mass term $\sim k^2$, but only for momenta $q^2 \lesssim k^2$. Correspondingly, the propagator $G_k$ in the presence of the IR cutoff replaces in \cref{eqn:propagator} $q^2 \to q^2 + R_k(q)$. We will start the flow for large enough $k$ such that the IR cutoff prevents any singular behavior of $G_k$, e.g. $k^2 \gg |m^2|$. As $k$ is lowered, one gradually approaches the potential singular behavior.

In a flat background $\tensor{t}{_\mu_\nu}(q)$ is an irreducible tensor representation of the rotation group $SO(4)$ (or Lorentz group for Minkowski signature). It therefore cannot mix on the quadratic level with scalar or vector representations. The transversal traceless fluctuations are physical metric fluctuations that do not mix with the gauge fluctuations \cite{wetterich2016quantum}. These properties make the flow equation \labelcref{eqn:wetterich} block-diagonal and allow us to deal with the $\tensor{t}{_\mu_\nu}$-contribution to the flow separately. Additional pieces in the flow equation from metric fluctuations beyond $\tensor{t}{_\mu_\nu}$, gauge fixing terms and ghosts \cite{reuter1994effective,reuter1998nonperturbative} will be neglected for our discussion. They may partly cancel contributions from fluctuations in $\tensor{t}{_\mu_\nu}$, without changing the overall picture. With respect to the rotation group $SO(3)$ the transversal traceless fluctuations contain the two degrees of freedom of the graviton. They also account for two vector and one scalar degree of freedom that do not correspond to propagating particles if the background obeys the field equations \cite{wetterich2016quantum}. In a later part of this note we will focus on the propagating graviton fluctuations. This will not affect the main findings. In the first part on flat space we employ ``graviton'' in the wider sense of the full irreducible representation $\tensor{t}{_\mu_\nu}$.

We are interested in the flow of the cosmological constant $V$. It is obtained directly by evaluating \cref{eqn:wetterich} for vanishing metric fluctuations, $\tensor{g}{_\mu_\nu} = \tensor{\delta}{_\mu_\nu}$, such that $\Gamma = \int_x V$ results in the simple one-loop expression
\begin{equation}\label{eqn:cc flow}
    \partial_t V
    = k \, \partial_k V
    = 5 I_k\bigl(-\tfrac{2 V}{M^2}\bigr),
\end{equation}
with
\begin{equation}\label{eqn:loop integral}
    I_k(m^2)
    = \frac{1}{2} \int_q \bigl(q^2 + R_k(q) + m^2\bigr)^{-1} \partial_t R_k(q).
\end{equation}
Employing the Litim cutoff \cite{litim2001optimized}
\begin{equation}
    R_k(q)
    = (k^2 - q^2) \, \theta(k^2 - q^2)
\end{equation}
replaces $q^2 + R_k \to k^2$ for $q^2 < k^2$, while not affecting the propagator $(q^2 + m^2)^{-1}$ for $q^2 > k^2$. It permits to solve the momentum integral analytically
\begin{equation}
    I_k(m^2)
    = \frac{1}{32 \pi^2} \frac{k^6}{k^2 + m^2}.
\end{equation}

The flow equation \labelcref{eqn:cc flow} for $V$ has a very intuitive interpretation: one takes the one-loop formula for the contribution of the graviton fluctuations $\sim \int_q \ln(q^2 - 2 V/M^2)$, replaces $q^2 \to k^2$ in the IR region, and takes a logarithmic $k$-derivative. We will see below that the same type of equation holds in Minkowski space, with $q^2 = -\omega^2 + \vec{q}^2$, $q^0 q_0 = -\omega^2$, in the momentum integration \labelcref{eqn:loop integral}.

The qualitative features of the solution of the flow equation are best discussed in terms of the dimensionless variable
\begin{equation}
    v
    = \frac{2 V}{M^2 k^2}.
\end{equation}
Approximating first $M$ by a constant, the flow equation reads
\begin{equation}\label{eqn:v flow}
    \partial_t v
    = \beta_v
    = - 2 v + \frac{5 k^2}{16 \pi^2 M^2} (1 - v)^{-1}.
\end{equation}
This is a central equation of this note. Its generalization to the flow of the effective potential for scalars is the basis for the proposed solution of the cosmological constant problem. We observe the pole at $v = 1$, which is multiplied by a small factor for $k^2 \ll M^2$. The behavior of $\beta_v$ is shown for two values of $k^2/M^2$ in \cref{fig:beta v}. As $k$ is lowered, the transition between the two regimes becomes very sharp. The singular second term in \cref{eqn:v flow} plays then a role only for $v$ very close to one.
\begin{center}
    \includegraphics[width=0.9\linewidth]{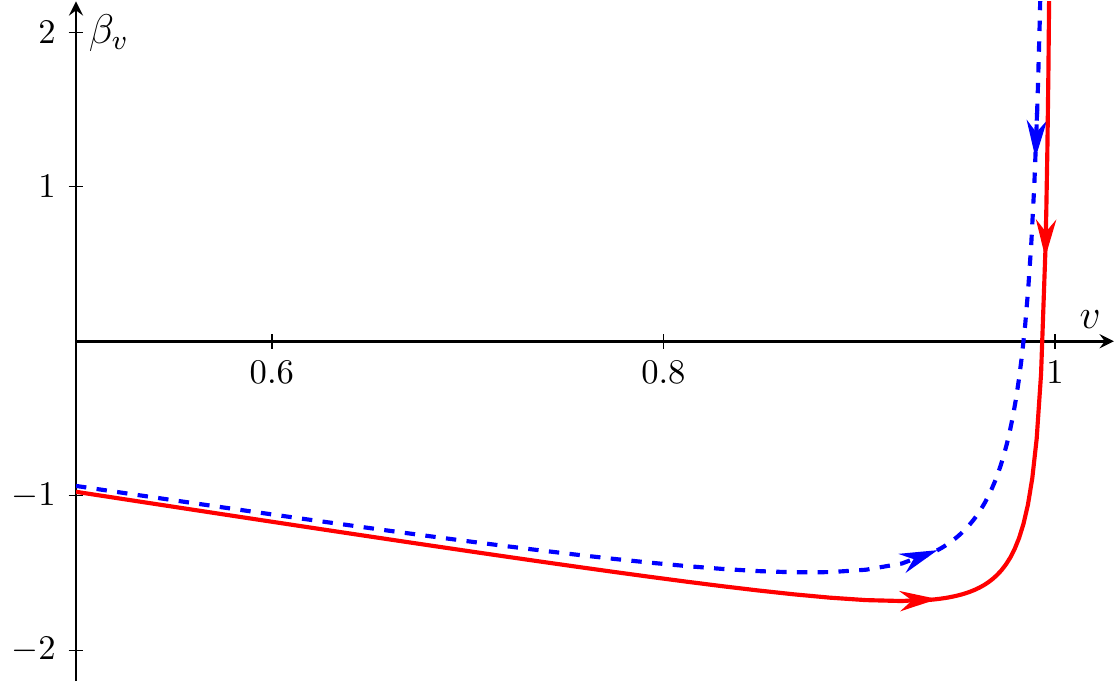}
    \captionof{figure}{Beta function for the dimensionless variable $v = 2 V/M^2 k^2$ for two values $k/M = 1$ and $k/M = 0.4$. The arrows denote the flow towards $k \to 0$ approaching the zero close to $v = 1$ (cf. \cref{eqn:v flow}).}
    \label{fig:beta v}
\end{center}

We concentrate on a positive cosmological constant, $v > 0$, and the range $v < 1$ for which the integral \labelcref{eqn:loop integral} is well defined. A small value of $v$ flows towards larger values according to $\beta_v \approx - 2 v$. On the other hand, values of $v$ very close to the pole at $v = 1$ decrease and $\beta_v$ is dominated by the positive second term. The solutions of the flow equation \labelcref{eqn:v flow} are attracted towards a sliding approximate fixed point $v_c(k)$ for which $\beta_v$ vanishes,
\begin{equation}
    v_c (1 - v_c)
    = \frac{5 k^2}{32 \pi^2 M^2}.
\end{equation}
This can be clearly seen by the numerical solutions of \cref{eqn:v flow} shown in \cref{fig:v}. This figure demonstrates in a simple way how initial values are ``forgotten''. The IR value $v = 1$ is reached independently of details of the initial conditions. In this way the cosmological constant problem will be solved without the need for fine-tuning of parameters.
\begin{center}
    \includegraphics[width=0.9\linewidth]{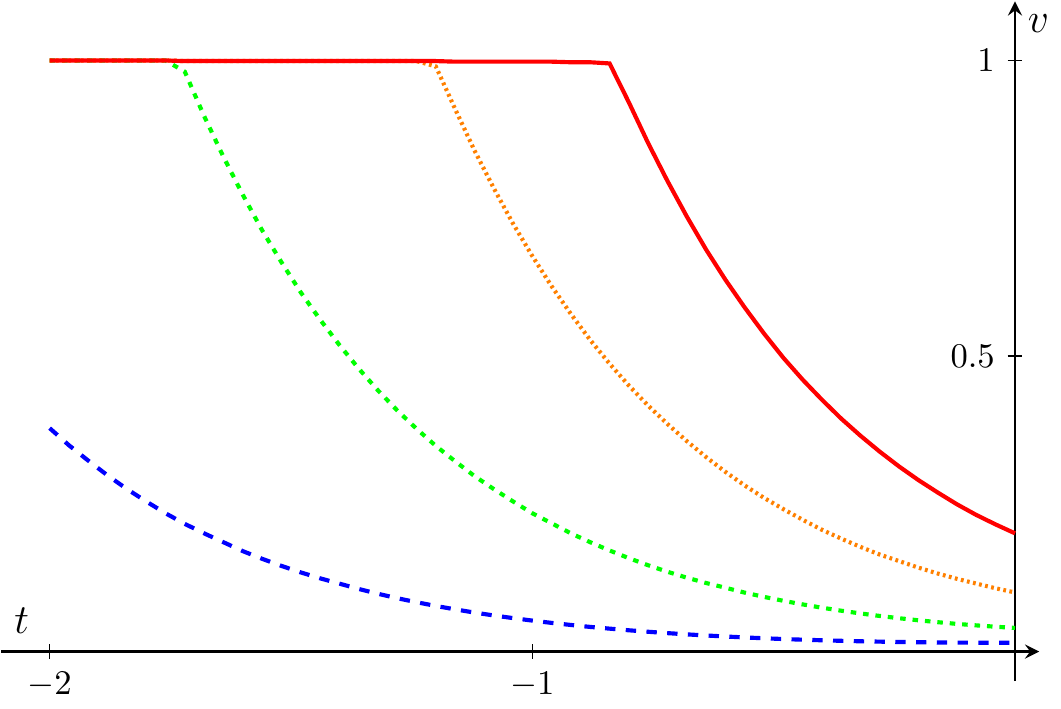}
    \captionof{figure}{Flow of the dimensionless variable $v = 2 V/M^2 k^2$ with $t = \ln(k/M)$ for different initial conditions. The trajectories shown all approach the quasi-fixed point $v_c \approx 1$ as $k$ is lowered.}
    \label{fig:v}
\end{center}

For $k^2 \ll 32 \pi^2 M^2/5$,  the relevant $v_c$ is close to 1, implying for the flow of the cosmological constant
\begin{equation}\label{eqn:ccc flow}
    V_c(k)
    = \frac{M^2 k^2}{2} \, v_c(k)
    = \frac{M^2 k^2}{2} - \frac{5 k^4}{64 \pi^2}.
\end{equation}
As $k$ goes to zero the cosmological constant approaches the pole at $v = 1$, but it never crosses it. The graviton-induced infrared flow of the cosmological constant drives it to zero! This holds actually even if the approximate fixed point $v_c(k)$ is not reached, which can happen for initial values of $v$ very close to zero or one. Since the flow is confined to the region $0 \leq v \leq 1$, the flowing cosmological constant always obeys $0 \leq V \leq M^2 k^2/2$.

We conclude from this simple discussion that graviton fluctuations can have strong effects on the running of the cosmological constant, even in the region $k^2 \ll M^2$. For $k^2 \gg 2 V/M^2$ the flow $\partial_t V \sim k^4$ would rapidly become insignificant as $k$ goes to zero. This changes dramatically for $k^2 \approx 2 V/M^2$. The pole in \cref{eqn:cc flow,eqn:v flow} can enhance the $\beta$-function by a large factor and prevents $V$ from remaining larger than $M^2 k^2/2$. The observation that a potential singularity in flow equations is avoided has been made earlier within functional renormalization. This feature explains the approach to convexity of the effective potential in scalar theories \cite{tetradis1992scale}. Rather than being just a formal mathematical construction, this approach to convexity due to fluctuations is a physical effect \cite{ringwald1990average}, needed, for example, for the quantitative understanding of spontaneous nucleation in first-order phase transitions \cite{strumia1999region}.

For quantum gravity singular features of the flow of the cosmological constant have been noted and discussed since early stages \cite{reuter1998nonperturbative,reuter2004quantum,reuter2009background,reuter2009conformal,christiansen2014fixed,christiansen2016global,biemans2017quantum,nagy2013critical}, while consequences for the physical value of the cosmological constant and cosmological consequences remained unclear. In this note we argue that the ``avoidance of instabilities'' by the flow governs directly the value of the observable cosmological constant and determines the features of dark energy in cosmology.

The length scale of the fluctuations driving the cosmological constant to zero can be far in the infrared. If we only include fluctuations with $q^2 > k^2$ the effective cosmological constant $V$ has a positive value $V \approx M^2 k^2/2$. With Minkowski signature, this would correspond to a de Sitter space with Hubble parameter $H^2 = k^2/6$. In other words, the fluctuations with wavelengths close to but somewhat smaller than the ``would-be horizon'' are needed in order to drive the cosmological constant to zero.

At this point, we should mention two important implicit assumptions underlying our computation of the flow of $V$. The first is that the field equations for the graviton, the graviton propagator, and its interactions can be derived from a diffeomorphism invariant effective action involving only one metric field. The second states that this effective action can be approximated by a derivative expansion to second order. These assumptions imply \cref{eqn:einstein hilbert action} at leading order. They are the basis of all computations in classical gravity and therefore well tested by experiment. Within functional renormalization it has been argued \cite{wetterich2016quantum,wetterich2016gaugeinvariant,wetterich2017gauge} that these properties indeed hold, provided one chooses a suitable physical gauge fixing or, equivalently, a constraint on conserved sources and physical fluctuations. These assumptions imply that the effective action for traceless transversal tensor fluctuations around flat space can be expanded as
\begin{equation}\label{eqn:diffeomorphism constraint}
    \Gamma
    = \int_x \sqrt{\bar{g}} \Bigl(V - \frac{V}{4} \tensor{t}{^\mu^\nu} \tensor{t}{_\mu_\nu} - \frac{M^2}{8} \tensor{t}{^\mu^\nu} \partial^2 \tensor{t}{_\mu_\nu} + \dots\Bigr)
\end{equation}
with $\sqrt{\bar{g}} = 1$ for Euclidean signature and $\sqrt{\bar{g}} = i$ for Minkowski signature, and $\partial^2 = \tensor{\bar{g}}{^\mu^\nu} \partial_\mu \partial_\nu$. The dots denote the graviton interactions. In the limit of vanishing momentum $(\partial_\rho \tensor{\gamma}{^m^n} = 0$), the graviton propagator and all interactions are determined by a single constant $V$, which is also the constant part in $\Gamma$. This is the ``diffeomorphism constraint'' on the effective action for the graviton. We employ this constraint and evaluate the flow of $V$ by the flow of the constant term.

In pure gravity a $k$-independent value of $M^2$ is typically relevant for the IR running. Close to an ultraviolet (UV) fixed point one expects instead a scaling behavior $M^2 = f k^2$ \cite{reuter1998nonperturbative}. Our discussion can easily be extended to $M^2$ depending explicitly on $k$. In this case, the factor $M^2$ in the cutoff \labelcref{eqn:ir cutoff} leads to the replacement $\partial_t R_k \to \partial_t R_k + (\partial_t \ln M^2) R_k$ in the integrand $J$. For $M^2 = f k^2$ one replaces \cref{eqn:v flow} by
\begin{equation}
    \partial_t v
    = -4 v + \frac{5}{16 \pi^2 f} (1 - v)^{-1}.
\end{equation}
(More precisely, the second term on the r.h.s.\ is multiplied by 4/3 if we replace in the IR cutoff \labelcref{eqn:ir cutoff} $M^2 \to f k^2$. Throughout this note, we will neglect terms from the $k$-dependence of $M^2$ in the cutoff \labelcref{eqn:ir cutoff}.) The corresponding flow shows two fixed points, determined by
\begin{equation}\label{eqn:fixed points}
    v_\ast (1 - v_\ast)
    = \frac{5}{64 \pi^2 f}.
\end{equation}
The solution with smaller $v_\ast$ corresponds to the UV fixed point, whereas the larger $v_\ast$ is IR-attractive. In the UV scaling region, one has $V \sim k^4$. For $M^2(k) = f k^2 + M^2$ the IR fixed point in \cref{eqn:fixed points} is no longer realized if $k^2 < M^2/f$, and the flow switches to the behavior \labelcref{eqn:v flow}. This reflects some of the qualitative features observed in more involved functional renormalization studies with a flat background geometry \cite{christiansen2016global}. (Note that the concept and definition of the cosmological constant in \cite{christiansen2016global} differs from the present one. The diffeomorphism constraint \labelcref{eqn:diffeomorphism constraint} is not realized in this work.)

\section{Flowing scalar potential}

Our simple discussion can be extended to the graviton contributions to the flow of the effective (average) potential $U_k(\chi)$ of some scalar field $\chi$. (See \cite{percacci2003asymptotic,narain2010renormalization,oda2016non,dona2016asymptotic,henz2017scaling,eichhorn2016quantum} for more extended quantum gravity investigations with scalar fields.) One simply replaces in the flow equation \labelcref{eqn:cc flow} $V$ by $U(\chi)$. Here $U(\chi)$ is defined by evaluating $\Gamma$ for vanishing metric fluctuations in flat space, and for constant $\chi$, $\partial_\mu \chi = 0$, i.e. $\Gamma = \int_x U(\chi)$. The scalar potential describes all scalar interactions at zero momentum. The graviton contribution does not change if we replace $V$ by $U(\chi)$. Flow equations for $\chi$-derivatives, such as the scalar mass term $\partial^2 U/\partial^2 \chi$, can be obtained by taking $\chi$-derivatives of the flow equation for $U(\chi)$. The implicit assumption underlying this framework states that the zero-momentum limit for the field equations, propagators and interactions of the physical graviton and scalar field fluctuations can be derived from a diffeomorphism-invariant effective action $\Gamma = \int_x \sqrt{g} \, U(\chi)$.

In addition to the graviton contribution, there are other contributions to the flow of $U$. For the example of a canonical kinetic term for $\chi$, the combination of the fluctuations of the scalar $\chi$ and the graviton yields in a simple truncation
\begin{equation}\label{eqn:potential flow}
    \partial_t U
    = \frac{k^6}{32 \pi^2} \biggl(\frac{5}{k^2 - 2 U/M^2} + \frac{1}{k^2 + \partial^2 U/\partial^2 \chi}\biggr).
\end{equation}
The second term \cite{wetterich1993exact,wetterich2001effective} involves the $\chi$-dependent scalar mass term, as given by the second derivative of $U$. We assume here a range for which the mixing of fluctuations in $\chi$ with scalar fluctuations in the metric can be neglected.

We distinguish two regimes. As long as $k \gg |2 U/M^2|$, the first term mainly contributes to the flow of a $\chi$-independent additive constant. The flow of the $\chi$-dependence of $U$ is governed by the scalar fluctuations, with a negligible graviton contribution. This yields the standard flow of the effective average potential \cite{wetterich2001effective}. For models with a discrete symmetry $\chi \to -\chi$, the potential only depends on $\rho = \chi^2/2$, with $\partial^2 U/\partial^2 \chi = U^\prime + 2 \rho U^{\prime\prime}$ and primes denoting derivatives with respect to $\rho$. In a truncation where $U^{\prime\prime\prime}$ and higher $\rho$-derivatives are neglected, the flow of $U^\prime$ and $U^{\prime\prime}$ can be computed by taking $\rho$-derivatives of \cref{eqn:potential flow}, i.e.
\begin{equation}
    \partial_t U^\prime
    = -\frac{k^6}{32 \pi^2} \biggl\{\frac{3 U^{\prime\prime}}{(k^2 + U^\prime + 2 \rho U^{\prime\prime})^2} - \frac{10 U^\prime}{M^2 (k^2 - 2 U/M^2)^2}\biggr\}
\end{equation}
and
\begin{equation}\label{eqn:upp flow}
    \partial_t U^{\prime\prime}
    = \frac{k^6}{16 \pi^2} \biggl\{\frac{9 U^{\prime\prime2}}{(k^2 + U^\prime + 2 \rho U^{\prime\prime})^3} + \frac{5 U^{\prime\prime}}{M^2 (k^2 - 2 U/M^2)^2}\biggr\}.
\end{equation}
The graviton corrections are suppressed by a factor $1/M^2$. (In \cref{eqn:upp flow}, we keep only  the leading power.) If we identify the quartic scalar coupling $\lambda$ with $U^{\prime\prime}(\rho_0)$, with $\rho_0$ the location of the minimum of $U(\rho)$, and take $k^2 \gg |U^\prime + 2 \rho U^{\prime\prime}|$, $k^2 \gg |2 U/M^2|$, we recover the usual one-loop running up to a gravitational correction,
\begin{equation}\label{eqn:lambda flow}
    \partial_t \lambda
    = \frac{9 \lambda^2}{16 \pi^2} + \frac{5 \lambda k^2}{16 \pi^2 M^2}.
\end{equation}

For particle physics experiments, the external momenta of the scattered particles typically set an effective IR cutoff. Identifying $k$ roughly with this cutoff, it will be in the \si{\giga\electronvolt} range. (A few orders of magnitude play no role here.) For a potential
\begin{equation}\label{eqn:higgs potential}
    U_\text{SM}
    = V + \frac{\lambda}{2} (\rho - \rho_0)^2
\end{equation}
mimicking the one of the Higgs scalar, gravity effects are negligible in the range $\lambda \rho^2 \ll k^2 M^2$. Similar features hold for extended models where Yukawa couplings to fermions or gauge interactions are taken into account.

The second regime concerns the ``singular region'' $k^2 \approx 2 U/M^2$. For a potential of the form \labelcref{eqn:higgs potential}, this always becomes relevant for very large $\rho$, namely $\rho \approx k M/\sqrt{\lambda}$. In this region, the graviton-induced renormalization effects become strong and the polynomial form of $U$ is no longer maintained. Typically, the potential flattens for very large $\rho$, such that $U(\rho) \lesssim M^2 k^2/2$ and the pole will not be crossed. A good approximation to the result of the flow is given by
\begin{equation}\label{eqn:potential approx}
    U(\rho)
    = \begin{cases*}
        U_\text{SM}(\rho) & for $\rho < \rho_c$,\\
        M^2 k^2/2 & for $\rho > \rho_c$,
    \end{cases*}
\end{equation}
with $\rho_c$ dependent on $k$  according to
\begin{equation}
    U_\text{SM}(\rho_c)
    = \frac{M^2 k^2}{2}.
\end{equation}
The transition region for $\rho$ near $\rho_c$ will be smoothened, but the behavior for $\rho \gg \rho_c$ is given to a good accuracy by the flat part of the potential \labelcref{eqn:potential approx}.

As $k$ is lowered, $\rho_c(k)$ decreases and the field region where the ``standard model'' potential $U_\text{SM}$ is valid shrinks. For $V < 0$ in \cref{eqn:higgs potential}, the standard model region remains finite with $\smash{\bigl(\rho_c(k=0) - \rho_0\bigr)^2} = -2 V/\lambda$. For $V > 0$, however, the  potential becomes entirely flat for small enough $k$, given by the $k$-dependent cosmological constant discussed previously. While the range of $\rho$ where the flattening happens is not relevant for particle scattering, it may be very important for cosmology and the cosmological constant problem.

We can also extend our simplified discussion to the scaling regime near a UV fixed point, where $M^2 = f k^2$. This results for \cref{eqn:lambda flow} in the gravity-induced anomalous dimension $A_\lambda$,
\begin{equation}
    \partial_t \lambda
    = A_\lambda \lambda + \frac{9 \lambda^2}{16 \pi^2},
    \qquad
    A_\lambda
    = \frac{5}{16 \pi^2 f}
    = \frac{5 g_\ast}{2 \pi}.
\end{equation}
The anomalous dimension is positive, $A_\lambda > 0$, such that $\lambda$ is driven towards zero as $k$ is lowered. Typically, $A_\lambda$ is of the order one, such that the approach to $\lambda = 0$ is fast. Values found for the scaling gravitational coupling in functional renormalization group investigations of gravity are in the range $g_\ast = G_\text{N} k^2 = (8 \pi f)^{-1} = 0.3$ \cite{souma1999non}, 1.4 \cite{litim2004fixed}, 1.0 \cite{manrique2011bimetric}, 2.0 \cite{christiansen2014fixed}, 1.5 \cite{codello2014consistent}, 0.8\,-\,1.8 \cite{dona2014matter}, 0.7\,-\,0.9 \cite{gies2015generalized}, 0.8 \cite{christiansen2016global}, 1.2 \cite{oda2016non}.

\section{Cosmon potential in variable gravity and the solution of the cosmological constant problem}

In general, $M^2$ is a function of the ``cosmon'' scalar field $\chi$. For large $\chi$, one expects $M^2 \sim \chi^2$, and we will choose a normalization of the scalar field where $M = \chi$. Adding a kinetic term for $\chi$,
\begin{equation}\label{eqn:vg effective action}
    \Gamma
    = \int_x \sqrt{g} \biggl\{-\frac{M^2(\chi)}{2} \, \tilde{R} + U(\chi) + \frac{1}{2} K(\chi) \partial^\mu \chi \partial_\mu \chi\biggr\}
\end{equation}
constitutes the effective action for ``variable gravity'' \cite{wetterich2014variable}.(Often $M^2(\chi)$ is denoted by $F(\chi)$ and $U(\chi)$ by $V(\chi)$.)

For $M^2 = \chi^2$ the graviton contribution to the flow of $U$ takes the form
\begin{equation}\label{eqn:graviton contribution}
    \partial_t U
    = \frac{5 k^6}{32 \pi^2} (k^2 - 2 U/\chi^2)^{-1} + \dots,
\end{equation}
where the dots denote contributions from the scalar fluctuations and other degrees of freedom. In the region of $\chi$ where the graviton contribution dominates, we can take over the discussion of the running cosmological constant, and \cref{eqn:ccc flow} turns for $k^2 \ll \chi^2$ to
\begin{equation}\label{eqn:large chi u}
    U
    = \frac{k^2}{2} \, \chi^2.
\end{equation}
This region corresponds to the close vicinity of the pole in \cref{eqn:v flow} and applies to the asymptotic behavior for large $\chi$. The cosmon potential increases for large $\chi$ quadratically, while a term $U \sim \chi^4$ would lead to a crossing of the pole and is therefore not compatible with the flow of $U$. Strong gravity-induced renormalization effects prevent an increase of $U$ (for increasing $\chi$) faster than $\chi^2$.

The IR potential \labelcref{eqn:large chi u} is universal. Due to the strong attraction to the fixed point it is independent of the initial conditions of the flow - no fine tuning of parameters is required. We will discuss below that the potential \labelcref{eqn:large chi u} is also independent of the choice of the cutoff function $R_k(q)$, up to a multiplicative constant $\bar{k}/k$ in the definition of $k$. For fixed $k$ the IR potential governs the asymptotic behavior for large $\chi$. For a more general dependence of $M$ on $\chi$ the universal asymptotic form of the potential for $\chi \to \infty$ becomes
\begin{equation}
    U
    = \frac{\bar{k}^2}{2} M^2(\chi).
\end{equation}
This is the central ingredient for the proposed solution of the cosmological constant problem.

For a scaling solution the dimensionless functions characterizing $\tilde{L}$ in $\Gamma = k^4 \int_x \tilde{L}$ depend only on dimensionless combinations such as $y = \chi^2/k^2$. In order to use the canonical dimension of mass for $\chi$ we have to select a particular $k$ that sets the units. We denote this as $\bar{k} = \sqrt{2} \mu$, such that for $M = \chi$ one has $U = \mu^2 \chi^2$. The value of $\mu$ is arbitrary and has no physical content - it only sets the units for $\chi$. The ratio $\chi/\mu$ can be used to denote the position on the flow trajectory corresponding to $\sqrt{2 y}$ for the scaling solution. Cosmological observables will only depend on $y$. For a fixed $\mu$ the IR limit corresponds to $\chi \to \infty$.

Variable gravity with quadratic cosmon potential has been studied extensively for cosmology \cite{wetterich2014variable,wetterich2015inflation,rubio2017emergent}. Depending on the precise form of the kinetic term, it can describe inflation for an early period where $\chi$ remains not too large, and dynamical dark energy for the region of large $\chi$. Indeed, the cosmological field equations are solved by $\chi$ increasing from small values towards infinity in the infinite future. At late time cosmology therefore explores the IR limit. If $\chi$ increases to infinity for $t \to \infty$, the infinite future corresponds to the IR fixed point.

This type of dynamics solves the cosmological constant problem and provides for a simple mechanism for dynamical dark energy. Indeed, the dimensionless effective cosmological constant vanishes asymptotically for large time \cite{wetterich1988cosmology}. It is given by the dimensionless ratio between the potential and the fourth power of the dynamical Planck mass,
\begin{equation}
    \frac{U}{M^4(\chi)}
    = \frac{\mu^2}{M^2(\chi)}
    = \frac{\mu^2}{\chi^2}
    \to 0.
\end{equation}
(The last identity refers to $M(\chi) = \chi$.) The effective cosmological constant vanishes asymptotically for all cosmological solutions for which the dynamical Planck mass $M(\chi)$ diverges in the infinite future. In the Einstein frame, this corresponds (approximately) to an exponential decrease of the potential as a function of a scalar field $\varphi$ with appropriately normalized kinetic term \cite{wetterich1988cosmology}.

Since in the present epoch $\chi$ is still finite, the effective cosmological constant does not yet vanish. It decreases, however, with time, yielding the first prediction of dynamical dark energy or quintessence \cite{wetterich1988cosmology}. What is crucial for this dynamical solution of the  cosmological constant problem is an increase of $U$ for large $\chi$ slower than $\sim M^4(\chi)$. In our present setting, this is enforced by the structure of the graviton contributions to the flow of $U$. The central result of this note simply states that for $M \sim \chi$ it is impossible that $U$ increases $\sim \chi^4$ for $\chi \to \infty$. Such an increase would inevitably lead to a strong instability of the graviton fluctuation effects, which is avoided by the flow of the renormalized couplings.

The contribution of scalar fluctuations omitted in \cref{eqn:graviton contribution} is somewhat more complicated than in \cref{eqn:potential flow}. This is due to mixing with the scalar degrees of freedom in the metric, as well as to a possible dependence of the kinetic coefficient $K$ on $\chi$. Near the pole, the graviton contributions will dominate, however, such that the scalar fluctuations will not modify the asymptotic behavior \labelcref{eqn:large chi u} for large $\chi$. This extends to other neglected fluctuation contributions as well.

We notice that the IR fixed point leading to \cref{eqn:large chi u} is not the only possible fixed point. An IR unstable fixed point can be obtained for $U(\chi \to \infty) \sim k^4$. This type of fixed point generalizes the solution of \cref{eqn:fixed points} for the smaller $v_\ast$, where the pole in the graviton contribution plays no role. Candidates for this type of scaling solution have been found in the discussion of dilaton quantum gravity in \cite{henz2017scaling}. Our simplified discussion suggests that perturbations of this type of scaling solutions are unstable and trigger a flow to the IR fixed point with the behavior \labelcref{eqn:large chi u}. In any case, a behavior $U \sim \mu^4$ would also solve the cosmological constant problem dynamically.

No matter what are the precise details in the variable gravity setting, the effective dynamical cosmological constant cannot be negative if $|U(\chi)|$ increases with some power of $\chi$. A negative value of the relevant asymptotic potential $U(\chi \to \infty)$ would lead to an instability in the scalar sector that is not compatible with a consistent quantum field theory. While bosonic fluctuations drive $U(\chi)$ to smaller values as $k$ is lowered, fermionic fluctuations tend to increase $U$. Near the pole in the graviton contribution, the bosonic graviton fluctuations always win and enforce the asymptotic behavior \labelcref{eqn:large chi u}. If bosonic fluctuations also win in the region of small $\chi$ they may lead to negative values of $U$ in this region. Nevertheless, the bosonic fluctuations cannot change the sign of the potential for large $\chi$ if the latter diverges for $\chi \to \infty$. Only this asymptotic behavior counts for the observable effective cosmological ``constant''. In summary, the strong IR fluctuation effects induced by the graviton predict a positive dynamical dark energy that vanishes in the asymptotic future.

\section{Graviton contributions to the Higgs potential}

The cosmon field is not the only scalar field, and one may wonder what happens to the renormalization flow of the effective potential in the presence of several scalar fields. We investigate this issue in the context of variable gravity. We concentrate on the Higgs doublet $h$, with straightforward generalization to other scalar fields, including composite scalars such as the chiral condensate in QCD. (Similar results are found in gravity coupled only to the Higgs boson. One replaces below $\chi$ by a fixed $M$.)

Let us assume that for $k$ much smaller than $\chi$, say $k = \SI{100}{\giga\electronvolt}$, we can approximate for a suitable field range of $h$ the potential by
\begin{equation}\label{eqn:potential approximation}
    \begin{aligned}
        &U
        = U_h(h,\chi) + \Delta U(\chi),\\
        &U_h
        = \frac{\lambda_h}{2} (h^\dagger h - \epsilon_h \chi^2)^2.
    \end{aligned}
\end{equation}
This range includes the partial minimum of $U$ with respect to $h$ which occurs for $h = (h_0,0)$, where $h_0 = \sqrt{\epsilon_\chi} \chi$ defines the Fermi scale. The Fermi scale is proportional to $\chi$ such that for constant Yukawa couplings and constant $\epsilon_h$ the ratio between the electron mass and Planck mass remains constant even for a cosmology with varying $\chi$. The quartic coupling of the Higgs scalar $\lambda_h(k,\chi)$ is of the order 1 for $k \sim \SI{100}{\giga\electronvolt}$, $\chi \sim \SI{e18}{\giga\electronvolt}$, while $\lambda_h \epsilon_h$ is in this region of $k$ and $\chi$ a tiny dimensionless coupling associated to the gauge hierarchy. The coefficient in front of the curvature scalar is given by the dynamical Planck mass that we take here as
\begin{equation}\label{eqn:planck mass}
    M^2
    = \chi^2 + \xi_h h^\dagger h.
\end{equation}

We want to study what happens if $k$ decreases further below \SI{100}{\giga\electronvolt}. For $h^\dagger h$ sufficiently close to $h_0^2$, the graviton fluctuations give only a negligible contribution to the flow of $U_h$. Fluctuations of particles with mass smaller than $k$, such as electrons or quarks, lead to a flow of $\lambda_h$ and $\epsilon_h$ according to the usual loop computation in particle physics. (The flow of $\lambda_h \epsilon_h$ is governed by an anomalous dimension \cite{wetterich1981gauge}.) This flow stops effectively for $k$ below the electron mass such that $h_0(k)/\chi$ reaches its final value. The graviton fluctuations contribute to the flow of $\Delta U(\chi)$, however. For $h = h_0$ the flow of $\Delta U$ is the same as the one for $U$ in \cref{eqn:graviton contribution} if we take for simplicity $\xi_h = 0$. In summary, the flow of $U_h$ is given by particle physics, while the flow of $\Delta U$ is strongly affected by the graviton fluctuations for the range of $\chi$ close to the pole in \cref{eqn:v flow}.

This simple picture holds, however, only for a range of $h^\dagger h$ sufficiently close to $h_0^2$. Since $U_h = \lambda_h \Delta^2/2$ increases quadratically with $\Delta = h^\dagger h - h_0^2$, graviton fluctuations become important at some critical $\Delta_c(k)$. The bound from the pole-like behavior reads now
\begin{equation}
    U_h + \Delta U
    \leq \frac{k^2}{2} (\chi^2 + \xi_h h^\dagger h).
\end{equation}
It will be saturated for large enough $\Delta$ such that the asymptotic behavior of $U_h$ is at most linear in $\Delta$.

We may estimate the critical value of $\Delta _c$ for the 
transition from the approximation \labelcref{eqn:potential approximation} to the behavior linear in $\Delta$ by
\begin{equation}
    \lambda_h \Delta_c^2
    = k^2 (\chi^2 + \xi_h \Delta_c),
    \qquad
    \Delta_c
    \approx k \chi.
\end{equation}
This value depends on $\chi$ and varies with $k$. For $k \approx \SI{100}{\giga\electronvolt}$, $\chi \approx \SI{e18}{\giga\electronvolt}$ one has $\Delta_c \approx (\SI{e10}{\giga\electronvolt})^2$, such that the particle physics potential \labelcref{eqn:potential approximation} can be trusted for $|h| \lesssim \SI{e10}{\giga\electronvolt}$. For particle physics experiments, the graviton fluctuations play no role. On present cosmological length scales, $k = \SI{e-33}{\electronvolt}$, the range of validity of the approximation \labelcref{eqn:potential approximation} shrinks to $\Delta < (\SI{e-3}{\electronvolt})^2$ or $|h - h_0| < \SI{e-17}{\electronvolt}$. This range seems tiny at first sight. We should, however, compare the energy density for an excitation $|h - h_0| = \SI{e-17}{\electronvolt}$, i.e. $\Delta \rho \sim h_0^3 |h - h_0| \sim (\SI{e4}{\electronvolt})^4$, with the present cosmological energy density $\rho_c \sim (\SI{2e-3}{\electronvolt})^4$. It is 28 orders of magnitude larger, which makes it clear that all relevant excitations of $h - h_0$ on cosmological scales are far below the critical amplitude of \SI{e-17}{\electronvolt}.

It is often argued that the change in the effective potential due to spontaneous electroweak symmetry breaking is a puzzle for a vanishing cosmological constant, since it contributes to $\Delta U$ an amount $\sim h_0^4 \sim (\SI{100}{\giga\electronvolt})^4$. It is interesting to see how our scenario of strong graviton-induced renormalization effects deals with this issue. At the scale $k$ relevant for electroweak symmetry breaking, $k \approx \SI{100}{\giga\electronvolt}$, the potential $\Delta U(\chi)$ still can reach values up to $k^2 \chi^2 \approx (\SI{e10}{\giga\electronvolt})^4$ for $\chi \sim \SI{e18}{\giga\electronvolt}$. Thus $h_0^4$ induces only a tiny change. As $k$ is lowered, $\Delta U$ follows its flow equation and reaches a value $h_0^4$ for $k \approx \SI{e-5}{\electronvolt}$. The graviton fluctuations with momenta smaller than \SI{e-5}{\electronvolt} finally renormalize the complete $\Delta U$ to even smaller values, absorbing in this way the jump due to electroweak symmetry breaking.

Beyond the IR regime we may approximate
\begin{equation}
    M^2
    = \chi^2 + f k^2.
\end{equation}
(As compared to \cref{eqn:planck mass} we have set $\xi_h = 0$.) In this approximation we can summarize the qualitative features of graviton contributions to the flow of the effective potential by the simple formula
\begin{equation}\label{eqn:simple formula}
    \partial_t U_g
    = \frac{5 k^6}{32 \pi^2} \biggl(k^2 - \frac{2 U}{\chi^2 + f k^2}\biggr)^{-1}.
\end{equation}
\Cref{eqn:simple formula} covers the whole range from the UV ($k \to \infty$) to the IR ($k \to 0$). With $\rho_h = h^\dagger h$ and primes denoting now derivatives with respect to $\rho_h$, one obtains the graviton contribution to the flow of derivatives of $U$ by taking corresponding derivatives of \cref{eqn:simple formula}, e.g.
\begin{equation}\label{eqn:graviton flow}
    \begin{aligned}
        &\partial_t U_g^\prime
        = A^{(g)} U^\prime,\\
        &A^{(g)}
        = \frac{5 k^6}{16 \pi^2 (\chi^2 + f k^2)} \biggl(k^2 - \frac{2 U}{\chi^2 + f k^2}\biggr)^{-2}.
    \end{aligned}
\end{equation}
Similarly, one finds at $U^\prime = 0$
\begin{equation}\label{eqn:quartic coupling flow}
    \partial_t U_g^{\prime\prime}\bigr|_{U^\prime = 0}
    = A^{(g)} U^{\prime\prime}.
\end{equation}

We identify the second derivative with respect to $\rho_h$ at the partial minimum of $U$ with respect to $\rho_h$ with a $k$- and $\chi$-dependent quartic self-interaction of the Higgs boson, $\lambda_h(k,\chi)$. According to \cref{eqn:quartic coupling flow} the graviton contribution to the flow of $\lambda_h$ is given by the positive anomalous dimension $A^{(g)}$. Depending on the value of $k/\chi$, we observe three regimes. For the UV regime $f k^2 \gg \chi^2$ one typically has $U = u k^4$ such that
\begin{equation}\label{eqn:anomalous dimension}
    A^{(g)}
    = \frac{5}{12 \pi^2 f \bigl(1 - v_\ast\bigr)^2},
    \qquad
    v_\ast
    = \frac{2 u}{f}.
\end{equation}
(We have included here the factor 4/3 from the full $k$-dependence of $R_k$.) With $8 \pi f = 1/g_\ast$ typically of the order one, the anomalous dimension is sizeable and not suppressed by any small parameter. Values $u/f$ (often called $\lambda$) vary in the literature, $u/f = 0.36$ \cite{souma1999non}, 0.26 \cite{litim2004fixed}, 0.22 \cite{manrique2011bimetric}, 0.22 \cite{christiansen2014fixed}, 0.0 \cite{codello2014consistent}, 0.1 \cite{dona2014matter}, 0.2 \cite{gies2015generalized}, 0.25 \cite{christiansen2016global}. These values, as well as the values for $g_\ast = (8 \pi f)^{-1}$ quoted above, are for pure gravity. They will change in the presence of matter particles. If we take for a rough estimate $u/f = 0.25$, $g_\ast = 1$, \cref{eqn:anomalous dimension} yields $A^{(g)} \approx 4$. Values of $A^{(g)}$ exceeding 2 seem well within the possibilities.

For the second regime one has $f k^2 \ll \chi^2$, while the flow is not yet dominated by the pole-like behavior such that $k^2 - 2 U/\chi^2$ is of the order $k^2$. As a consequence, $A^{(g)}$ is suppressed by the small ratio $k^2/\chi^2$. In this regime, the graviton contributions to the flow of $\lambda_h$ or $\lambda_h \epsilon_h$ are tiny. The flow of these quantities is dominated by the particle physics contributions, following the perturbative running in the standard model or extensions thereof. For $k^2 \ll m_e^2$, the flow eventually stops. Finally, the IR regime with strong gravitational infrared effects sets in when $k^2$ reaches the vicinity of $2 U/\chi^2$. The pole-like enhancement overwhelms the suppression $k^2/\chi^2$ and the anomalous dimension \labelcref{eqn:graviton flow} can get very large.

The large and positive anomalous dimension $A^{(g)}$ in the UV regime has important consequences for particle physics. First, $A^{(g)}$ drives $\lambda_h$ fast towards values close to zero. There are typically additional small particle physics contributions to the flow of $\lambda_h$ that do not necessarily vanish for $\lambda_h = 0$. Nevertheless, the large gravitational anomalous dimension dominates, resulting in a very small value of $\lambda_h$ at the scale $k = \chi/\sqrt{f}$ when the gravitational contributions die out. Subsequently, the contribution of the Yukawa coupling in the standard model lets $\lambda_h$ grow. This simple structure has led to the successful  prediction \cite{shaposhnikov2010asymptotic} of the value of the Higgs boson mass of \SI{126}{\giga\electronvolt} with a few \si{\giga\electronvolt} of uncertainty. The main requirement for this prediction, namely the gravity-induced positive and sizable anomalous dimension for the quartic scalar coupling $\lambda_h$, is realized within our simple computation.

For a second-order vacuum electroweak phase transition, the critical hypersurface cannot be crossed by the flow. The flow for a small  deviation from the critical hypersurface is governed by an anomalous dimension \cite{wetterich1981gauge,wetterich2016gauge}. With $\beta_{U^\prime} = \partial_t U^\prime$ the anomalous dimension reads $A = \partial \beta_{U^\prime}/\partial U^\prime\smash{\bigr|_\text{c.s.}}$, where the subscript reminds that $A$ has to be evaluated on the critical surface. If $A$ exceeds 2, the small ratio between Fermi scale and Planck mass $\sqrt{\epsilon_h}$ can be naturally explained by a ``resurgence mechanism'' \cite{wetterich2016gauge}. The dimensionless distance from the critical hypersurface $\gamma$ first shrinks to a very small value due to the flow in the UV regime. It then increases again, essentially due to the canonical dimension, once $k^2 < \chi^2/f$. Our computation yields a graviton contribution to $A$ which is given by $A^{(g)}$, cf. \cref{eqn:graviton flow}. Our simple estimate of its value \labelcref{eqn:anomalous dimension} in the UV regime suggests that it may indeed exceed 2. Establishing $A > 0$ requires, of course, a more complete quantum gravity computation. For a Higgs-Yukawa model coupled to gravity $A > 2$ has been found in ref.~\cite{oda2016non}.

\section{General instability-induced flow}

We have found strong quantum gravity effects in the infrared flow of the cosmological constant and the effective potential of the cosmon and the Higgs doublet. This perhaps surprising observation is related to an instability of the graviton propagator in flat space which occurs for (Euclidean) momenta $q^2 < 2 V/M^2$. Functional renormalization avoids this instability by a flow of $V$ towards zero. For momenta $q^2 \approx k^2$, the cosmological constant $V(k)$ or the scalar potential $U(\chi,k)$ is small enough for $2 V/M^2$ not to exceed $k^2$. (Here and in the following one can replace $V$ by $U$ and similarly $v$ by $2 U/M^2 k^2$.) This ``avoidance of the instability'' arises on the level of functional flow equations as a singular structure which prevents a flow into the unstable region. The effect is highly non-perturbative and will not be seen at any finite order of a perturbative expansion.

The consequences of a simple estimate of this strong infrared effect in quantum gravity are rather impressive:
\begin{enumerate}[label=\roman*.]
    \item The cosmological constant is renormalized to zero.
    \item Within variable gravity, dynamical dark energy is predicted.
\end{enumerate}
Extending the simple calculation of the graviton contribution to the UV regime and the functional flow of the Higgs potential leads to further results:
\begin{enumerate}[label=\roman*.,resume]
    \item A positive, sizable anomalous dimension for the quartic coupling of the Higgs doublet leads to a successful prediction of the mass of the Higgs boson.
    \item For a small enough flowing Planck mass in the UV scaling regime, the gauge hierarchy of the electroweak symmetry breaking can be explained.
\end{enumerate}
Needless to say that such dramatic consequences call for a critical assessment of the reliability of the simple estimate. We concentrate here on the IR regime and briefly address three questions:
\begin{enumerate}
    \item Is the strong IR gravity effect in Euclidean flat space robust with respect to a change of cutoff and a more complete treatment of the metric
    fluctuations?
    \item Does the result extend to Minkowski space?
    \item Is the strong IR effect also relevant if a realistic cosmological solution replaces flat space?
\end{enumerate}

On the technical level, the singular structure in the flow equation \labelcref{eqn:v flow} is the central ingredient. It turns $\beta_v$ necessarily positive if $v$ approaches one sufficiently closely. Since $v$ cannot grow beyond the value where $\beta_v$ turns positive, the singular behavior is avoided. For negative $m^2$ the divergence of $I_k(m^2)$ in \cref{eqn:loop integral} for a certain ratio $k^2/|m^2|$ is indeed rather genuine for `` admissible''  cutoff functions $R_k(q)$. The cutoff should decrease with decreasing $k$ such that $\partial_t R_k > 0$. Therefore $I_k(m^2)$ is positive as long as $P(q) = q^2 + R_k(q)$ is positive. If furthermore $I_k(m^2)$ diverges as a critical value $v_c$ for $v = 2 V/M^2 k^2$ is approached, one necessarily obtains $\beta_v > 0$ in the vicinity of the singularity for $v_c$, and $v < v_c$. Thus $v$ either decreases (for $\beta_v > 0$) or it cannot increase beyond the point where $\beta_v$ vanishes. In consequence, the singularity at $v_c$ can never be crossed by the flow.

For the strong IR gravity effect it is sufficient that $I_k(v)$ has a singularity for finite $v > 0$. Let us denote $x = q^2/k^2$ and $\smash{p(x) = \bigl(q^2 + R_k(q^2)\bigr)/k^2}$, such that
\begin{equation}
    I_k(v)
    = \frac{k^4}{32 \pi^2} \int_0^\infty \dif x \, x \bigl(p(x) - v\bigr)^{-1} (\partial_t R_k/k^2).
\end{equation}
We consider cutoff functions for which $p(x)$ has a minimum for some $\bar{x} \neq 0$, $p(\bar{x}) = \bar{p}$, and is analytic in this region,
\begin{equation}\label{eqn:analytic p}
    p(\bar{x})
    = \bar{p} + a(x - \bar{x})^2 + \dots
\end{equation}
The case $a = 0$, $\bar{p} = 1$ corresponds to the Litim cutoff discussed above, and we extend the discussion now to $a > 0$. The singularity occurs for $v_c = \bar{p}$. For $\epsilon = \bar{p} - v \to 0$, $(\partial_t R_k/k^2)(\bar{x}) = 2 \bar{s}$, one can approximate the dominant integration region for $x$ close to $\bar{x}$ by
\begin{equation}\label{eqn:dominant region}
    I_k(v)
    = \frac{k^4 \bar{x} \bar{s}}{16 \pi^2} \int \frac{\dif x}{\epsilon + a (x - \bar{x})^2}
    = \frac{k^4 \bar{x} \bar{s}}{16 \pi \sqrt{a \epsilon}}.
\end{equation}
This integral indeed diverges with $\epsilon^{-1/2}$ for $\epsilon \to 0$, thus establishing that for this type of cutoff functions the singularity cannot be crossed \cite{tetradis1992scale}. As is well known from the investigation of the approach to convexity for scalar effective potentials, cutoff functions with minimum $p(x)$ at $x = 0$ are less suited. The lack of a singularity in $I_k(v)$ would entail a rather complex IR behavior that cannot be described anymore by a simple truncation.

From \cref{eqn:dominant region} the flow near the singularity is approximated by
\begin{equation}
    \partial_t v
    = -2 v + \frac{10}{M^2 k^2} \, I_k(v)
    \approx \frac{\bar{e}}{M^2 k^2} (\bar{p} - v)^{-1/2} - 2 \bar{p},
\end{equation}
with
\begin{equation}
    \bar{e}
    = \frac{5 \bar{x} \bar{s}}{8 \pi \sqrt{a}}.
\end{equation}
The approximate solution reads
\begin{equation}\label{eqn:approximate solution}
    v
    = \bar{p} - \biggl(\frac{\bar{e} k^2}{2 \bar{p} M^2}\biggr)^2,
    \qquad
    V
    = \frac{\bar{p} M^2 k^2}{2} - \frac{\bar{e}^2 k^6}{8 \bar{p}^2 M^2}.
\end{equation}
The asymptotic value $V_c = M^2 \bar{k}^2/2$, $\bar{k}^2 = \bar{p} k^2$, is approached even closer than in \cref{eqn:ccc flow}. For admissible cutoffs the behavior \labelcref{eqn:approximate solution} seems at first sight more generic than the limiting case of the Litim cutoff. We observe, however, that for small $\epsilon$ only a tiny range $|x - \bar{x}| \sim \sqrt{\epsilon/a}$ contributes substantially to the integral \labelcref{eqn:dominant region}, while for the Litim cutoff the weight is more equally distributed in the range $x < 1$. The precise approach to the universal IR value $V = M^2 \bar{k}^2/2$ depends strongly on the cutoff. In contrast, the universal form $V = M^2 \bar{k}^2/2$ is independent of the choice of the cutoff up to a multiplicative constant $\bar{k}/k$ related to the precise definition of $k$. Only the universal IR value of $V$ is important for our purposes.

What about additional contributions to $\partial_t V$ from photons or other massless particles? Such fluctuations add to the flow of $V$ a term
\begin{equation}\label{eqn:massless fluctuations}
    \Delta \partial_t V
    = d k^4.
\end{equation}
This contribution is usually associated to a pledged ``unnaturalness of a small cosmological constant due to quantum or vacuum fluctuations''. In the presence of the strong IR graviton fluctuations an additional term \labelcref{eqn:massless fluctuations} in the flow of $V$ has almost no influence in this region of the flow. It modifies $V$ in \cref{eqn:approximate solution} only by a term $\sim k^8/M^4$. For the Litim cutoff it adds to $\beta_v$ in \cref{eqn:v flow} a term $2 d k^2/M^2$ which is strongly suppressed for small $k^2/M^2$. As a result, the last term in \cref{eqn:ccc flow} is divided by a factor $(1 - d k^2/M^2)$. The correction $\sim d k^6/M^2$ is of the same order as other subleading terms in the expansion of $V$. We conclude that the effect of ``vacuum fluctuations'' of virtual particles is well present and of the order expected from simple estimates. It is, however, overwhelmed by the graviton fluctuations and cannot impede the approach of $V$ towards zero as $k$ is lowered. Fluctuations of particles with mass $m_p$ are further suppressed by a factor $k^2/m_p^2$ in the range $k \ll m_p$.

So far, we have only included the contribution from the graviton fluctuations. We have paid little attention to issues such as gauge invariance, gauge fixing, other components of metric fluctuations, ghosts, or fluctuations of the cosmon $\chi$ or Higgs doublet $h$. The strong IR quantum gravity effects are related to a singular behavior of flow equations as an unstable region is approached. For the ``avoidance of instabilities'' we can neglect fluctuations which yield regular contributions in the field- and $k$-region relevant for the ``singular flow''. This typically happens for matter fluctuations. (Additional singularities in the scalar sector may arise for non-convex parts of the potential. They are well understood \cite{tetradis1992scale} and will not be considered here.) The trace part of the physical metric fluctuations has a propagator $\smash{G_k \sim \bigl(P(q) - V/(2 M^2)\bigr)^{-1}}$. For $P(q) \geq 2 V/M^2$, this potential singularity is not reached by the flow.

The propagator of the gauge fluctuations in the metric is determined mainly by the gauge fixing term. If the gauge fixing parameter $\alpha$ goes to zero, the inverse propagator $\sim (q^2/\alpha + \text{finite pieces})$ does not lead to a singular behavior in the relevant range. Also mixing between physical fluctuations and gauge fluctuations becomes negligible, cf. ref.~\cite{wetterich2017gauge}
for details. Finally, ghost contributions are not singular either. We conclude that the strong IR flow near the transition to instability is completely governed by the fluctuations in $\tensor{t}{_\mu_\nu}$. Of course, all other fluctuations contribute away from the singularity. They play a role for a precision estimate of the anomalous dimensions for the Higgs potential in the UV scaling regime.

\section{Potential flow in Minkowski space}

We next turn to the flow equation in Minkowski space with metric $\tensor{\bar{g}}{_\mu_\nu} = \tensor{\eta}{_\mu_\nu}$. The factor $\sqrt{g}$ in \cref{eqn:einstein hilbert action} equals $i$, thus modifying \cref{eqn:inverse propagator},
\begin{equation}
    \Gamma^{(2)}
    = i \biggl(\frac{M^2q^2}{4} - \frac{V}{2}\biggr) P^{(t)},
\end{equation}
with $q^2 = q^\mu q_\mu = \vec{q}^2 - \omega^2 e^{2 i \epsilon}$, $q_\mu = (-\omega,\vec{q})$, and $\epsilon \to 0_+$ indicating the path taken in $\omega$-integrations. Correspondingly, we also multiply the cutoff function $R_k$ by a factor $i$, and $G$ picks up a factor $-i$. The exact flow equation \labelcref{eqn:wetterich} holds for arbitrary signature such that
\begin{equation}
    \partial_t (iV)
    = \frac{1}{2} \Tr \int_q \partial_t R_k \, G,
\end{equation}
with $\int_q = (2 \pi)^{-4} \int \dif \omega \int \dif^3 q$ now performed with Minkowski signature and $\Tr$ the trace over internal indices. The contribution of the graviton fluctuations reads
\begin{equation}\label{eqn:minkowski graviton contribution}
    \begin{aligned}
        \partial_t V
        &= -5 i \tilde{I}_k,\\
        \tilde{I}_k
        &= \frac{1}{2} \int_q J(q)
        = \frac{1}{2} \int_{\vec{q}} \int_\omega J(\omega,\vec{q}),\\
        J(q)
        &= \biggl(q^2 + R_k(q) - \frac{2 V}{M^2}\biggr)^{-1} \partial_t R_k(q).
    \end{aligned}
\end{equation}

The $\omega$-integral can be considered as an integral along the real axis in the complex plane, $\omega = \omega_R + i \omega_I$. With Euclidean momentum $q_0 = -\omega_I$ and assuming that $J$ vanishes sufficiently fast for $|\omega| \to \infty$, one has
\begin{equation}
    \tilde{I}_k
    = i I_k - \Delta_1 + \Delta_3,
\end{equation}
with Euclidean integral $I_k$ given by \cref{eqn:loop integral}, and $\Delta_1$, $\Delta_3$ the clockwise contour integrals around the regions I and III. We define the regions as I: $\omega_R > 0$, $\omega_I > 0$, II: $\omega_R > 0$, $\omega_I < 0$, III: $\omega_R < 0$, $\omega_I < 0$, and IV: $\omega_R < 0$, $\omega_I > 0$. If $J$ is analytic in the regions I and III, the flow equation for $V$ is the same for Minkowski space and Euclidean flat space. If not, there are additional contributions $\sim 2 i (\Delta_1 - \Delta_3)$ that may contain an imaginary part.

Let us consider a situation where $J$ is analytic in regions I and III except for possible poles. With $R_k$ depending only on $q^2$ the integrand $J(\omega,\vec{q})$ only depends on $\omega^2$. Possible poles come in pairs $\pm \bar{\omega}_j(\vec{q})$. Near a pair of poles $j$ one has
\begin{equation}
    J
    = \frac{r_j(\omega^2)}{\omega^2 - \bar{\omega}_j^2}
    = \frac{r_j(\bar{\omega}_j^2)}{2 \bar{\omega}_j} \biggl(\frac{1}{\omega - \bar{\omega}_j} - \frac{1}{\omega + \bar{\omega}_j}\biggr).
\end{equation}
The residua in regions I and III have therefore opposite signs, $\Delta_3 = -\Delta_1$. As a result one finds
\begin{equation}
    \tilde{I}_k
    = i\Bigl(I_k + \sum_j K_j\Bigr),
    \qquad
    K_j
    = \int_{\vec{q}} \frac{r_j(\bar{\omega}_j^2)}{2 \bar{\omega}_j},
\end{equation}
with $j$ the sum over poles of $J$ in region I, located at $\bar{\omega}_j(\vec{q})$. For $v < 1$ and suitable cutoff functions the integrals $K_j$ do not show a singular behavior if $r_j(\bar{\omega}_j^2)/\bar{\omega}_j$ remains finite in the whole integration region. The integrands should fall off fast enough for large $|\vec{q}|$ due to the factor $\partial_t R_k$. The effect on the flow of any ``finite contribution'' in $\tilde{I}_k$ is suppressed for small $k$ by a factor $k^2/M^2$. In consequence, the singular structure in $\tilde{I}_k$ for $\bar{k}^2 - 2 V/M^2 \to 0$ arises from the Euclidean integral $I_k$. The ``avoidance of the singularity'' is the same for Minkowski and Euclidean signature.

We may map the regions in the complex variable $x = q^2/k^2 = x_R + i x_I$ onto the complex $\omega$-plane, recalling that a given $x$ corresponds to two values of $\omega$ with opposite sign. With our $i \epsilon$-definition of $q^2$ one finds that $x_I < 0$ maps into the $\omega$-regions I, III, while $x_I > 0$ corresponds to regions II, IV. For $x_I = 0$ the real values of $q^2$ are mapped to regions II, IV if $\vec{q}^2/k^2 > x_R$, while they belong to regions I, III for $\vec{q}^2/k^2 < x_R$. If $R_k(x)$ is a real function of $x$ this also holds for $J(x)$. Any pole of $J(x)$ at $z$ implies therefore the existence of another pole at $z^\ast$. Poles in the regions I, III can therefore only by avoided for all $\vec{q}^2$ if they occur all for real negative $x$. An ``ideal cutoff function'' $R_k$ would be such that $J(x)$ remains analytic in the regions I, III.

Infrared cutoff functions for Minkowski signature have been discussed by Floerchinger \cite{floerchinger2012analytic}, see also \cite{pawlowski2015real,kamikado2014real,strodthoff2017self}. We may consider an algebraic cutoff of the form
\begin{equation}\label{eqn:cutoff}
    R_k(q)
    = b k^2 \biggl(\frac{k^2}{k^2 + c q^2}\biggr)^n,
\end{equation}
with positive real constants $b$ and $c$. For $|\omega| \to \infty$ one has $J \sim |\omega|^{-2(n+1)}$ such that contours can indeed be closed at $|\omega| \to \infty$. The poles of $\partial_t R_k$ are in regions II and IV. If $z = z_R + i z_I$ denotes one of the zeros of the polynomial $S(x)$, $x = q^2/k^2$,
\begin{equation}
    S(x)
    = (x - v) (1 + c x)^n + b,
\end{equation}
e.g. $S(z) = 0$, poles are present in the regions I, III unless all zeros occur for real negative $z$. Typically, $S(x)$ has zeros away from the real axis such that the choice \labelcref{eqn:cutoff} does not correspond to an ``ideal cutoff''. Since the contributions from $\Delta_1 - \Delta_3$ are subleading we concentrate on the Euclidean momentum integral for this cutoff. For $n b c > 1$, the minimum of $p(x) = x + R_k/k^2$ occurs at $\bar{x} > 0$. We may choose $b$ such that $\bar{p} = p(\bar{x}) = 1$, $\bar{k} = k$. The integral $I_k$ has a singularity $\sim 1/\sqrt{\epsilon}$ which will prevent the flow from entering the singular region $v > 1$, leading to the same strong graviton-induced renormalization effects for $V$ as for Euclidean flat space.

Our choice of $b$ corresponds to
\begin{equation}
    \bar{x}
    = 1 - \tfrac{1}{n+1} \bigl(1 + \tfrac{1}{c}\bigr),
    \qquad
    b
    = \tfrac{1}{n c} (1 + c \bar{x})^{n+1},
\end{equation}
where $c > \tfrac{1}{n}$, $b > 1$. An example is $n = 2$, $c = 1$, $\bar{x} = 1/3$, $p(0) = b = 32/27$, $p(1) = 35/27$. In the region $0 < x < 1$ one finds only a small enhancement of $p(x)$ as compared to the Litim cutoff $p(x) = 1$. With
\begin{equation}\label{eqn:algebraic cutoff}
    \begin{aligned}
        p(x)
        &= x + b (1 + c x)^{-n},\\
        s(x)
        &= \frac{\partial_k R_k}{2 k^2}
        = b (1 + c x)^{-n} \bigl(n + 1 - \tfrac{n}{1 + c x}\bigr),
    \end{aligned}
\end{equation}
the Euclidean integrand
\begin{equation}
    J(x)
    = \frac{2 s(x)}{p(x) - v}
\end{equation}
hardly differs from the Litim cutoff
\begin{equation}
    J_\text{L}(x)
    = \frac{2}{1 - v} \theta(1 - x).
\end{equation}
Our Euclidean estimate of the flow of $V$ with a Litim cutoff therefore yields a valid approximation for the flow in Minkowski space with cutoff \labelcref{eqn:algebraic cutoff}, except very near the singularity. The singular behavior is of the type \labelcref{eqn:analytic p,eqn:dominant region}.

\section{Flow on cosmological backgrounds}

The flow in a flat background metric clearly shows the strong IR gravity effect. One would like to extend the discussion to fluctuations in the vicinity of a realistic cosmological solution, with fixed homogeneous and isotropic background metric $\tensor{\bar{g}}{_\mu_\nu} = a^2(\eta) \tensor{\eta}{_\mu_\nu}$ and conformal time $\eta$. In variable gravity, this is accompanied by a fixed background solution for the scalar field $\bar{\chi}(\eta)$. For the graviton fluctuations, the background metric enters the flow equation by replacing in $\Gamma^{(2)}$ and $R_k$ the dependence on $q^2$ by the covariant Laplacian, $q^2 \to -D^\mu D_\mu$, as well as by additional geometric terms in $\Gamma^{(2)}$ arising from a non-vanishing curvature scalar, for details see ref.~\cite{wetterich2016quantum}. In variable gravity, the background scalar field $\bar{\chi}(\eta)$ appears in the factor $M^2(\bar{\chi})$ in the cutoff term. This differs from $M^2(\chi)$ in $\Gamma^{(2)}$, which is evaluated for arbitrary $\chi$. As for our simplified discussion in flat space, we omit the difference between $\bar{\chi}$ and $\chi$ and evaluate the effective action for $\bar{\chi} = \chi$.

One is interested in the form of the effective average action $\Gamma_k$ in the vicinity of a realistic cosmological solution $a(\eta)$ and $\bar{\chi}(\eta)$. For cosmology, this will be sufficient to derive quantum field equations and the power spectrum of primordial fluctuations \cite{wetterich2015cosmic}. Similarly to simple scalar models, where the best results of simple truncations are obtained by expanding around the ground state solution (e.g. the minimum of the effective potential), we expect for quantum gravity the most reliable results for expansions around the cosmological background. While $\Gamma_k$ is a functional of arbitrary metrics $\tensor{g}{_\mu_\nu}(x)$ and scalar fields $\chi(x)$, the background metric $\tensor{\bar{g}}{_\mu_\nu}$ and scalar field $\bar{\chi}$ may be considered fixed. (For alternatives cf. ref.~\cite{wetterich2016gaugeinvariant}.) An optimal procedure adjusts $\tensor{\bar{g}}{_\mu_\nu}$ and $\bar{\chi}$ a posteriori to a solution of the field equations derived from the effective action for $k = 0$ (or $k = k_0$).

For the discussion on a curved background we concentrate on the graviton fluctuations which correspond to the traceless transversal tensor with respect to the rotation group $SO(3)$. In a homogeneous isotropic background the graviton can be identified \cite{wetterich2016quantum} with the components of $\tensor{t}{_\mu_\nu}$ in the three ``space directions'' $m,n \in \{1,2,3\}$ with constraints
\begin{equation}
    \tensor{t}{_m_n}
    = a^2 \tensor{\gamma}{_m_n} + \dots,
    \quad
    q^n \tensor{\gamma}{_m_n}
    = 0,
    \quad
    \tensor{\gamma}{^m_m}
    = 0.
\end{equation}
The projector $P^{(\gamma)}$ onto the graviton mode reads
\begin{equation}\label{eqn:space projector}
    \begin{aligned}
        &\tensor*{P}{^{(\gamma)}_m_n_p_q}
        = \frac{1}{2} \bigl(\tensor{Q}{_m_p} \tensor{Q}{_n_q} + \tensor{Q}{_m_q} \tensor{Q}{_n_p} - \tensor{Q}{_m_n} \tensor{Q}{_p_q}\bigr),\\
        &\tensor{Q}{_m^n}
        = \tensor{\delta}{_m^n} - \frac{q_m q^n}{q^2},
        \qquad
        \Tr P^{(\gamma)}
        = \tensor*{P}{^{(\gamma)}_m_n^m^n}
        = 2,
    \end{aligned}
\end{equation}
with $P^{(\gamma)} = 0$ if one index equals zero. In flat space a restriction to the graviton mode replaces effectively $P^{(t)}$ by $P^{(\gamma)}$. Thus the graviton part in the flow equation \labelcref{eqn:cc flow} replaces the factor 5 by 2. Restricting the flat space discussion to the contribution of fluctuations of the ``propagating graviton'' $\tensor{\gamma}{_m_n}$ multiplies the graviton contribution by a factor 2/5. This quantitative modification does not change any of our qualitative findings.

For the effective action \labelcref{eqn:vg effective action} the inverse graviton propagator becomes \cite{wetterich2016primordial}
\begin{equation}\label{eqn:inverse graviton propagator}
    \tilde{\Gamma}_\gamma^{(2)}
    = \frac{i}{4} \Bigl[A^2 \bigl(\hat{D} + R_k(\hat{D})\bigr) + \Delta_\gamma\Bigr] P^{(\gamma)},
\end{equation}
with
\begin{equation}
    \begin{aligned}
        &\hat{D}
        = \partial_\eta^2 + 2 \hat{\mathcal{H}} \partial_\eta + \vec{q}^2,
        \qquad
        \vec{q}^2
        = \tensor{\delta}{^m^n} q_m q_n,\\
        \label{eqn:delta}
        &\Delta_\gamma
        = -2 A^4 \hat{V} + 2 A^2 (\hat{\mathcal{H}}^2 + 2 \partial_\eta \hat{\mathcal{H}}) + A^2 \hat{K} (\partial_\eta \bar{\chi})^2.
    \end{aligned}
\end{equation}
Here we employ frame-invariant quantities, for $M = M(\chi)$, $U = U(\chi)$,
\begin{equation}\label{eqn:frame-invariant}
    \begin{aligned}
        &A
        = M a,
        \qquad
        \hat{\mathcal{H}}
        = \partial_\eta \ln A,\\
        &\hat{V}
        = \frac{U}{M^4},
        \qquad
        \hat{K}
        = \frac{K}{M^2} + \frac{3}{2 M^4} \biggl(\frac{\partial M^2}{\partial \chi}\biggr)^2.
    \end{aligned}
\end{equation}
We work in three-dimensional Fourier space with spacelike comoving momenta $q_m$, and in position space for conformal time $\eta$, $\eta^\prime$. The graviton Green's function obeys
\begin{equation}\label{eqn:identity}
    \tilde{\Gamma}_\gamma^{(2)} G_\gamma(\eta,\eta^\prime)
    = P^{(\gamma)} \delta(\eta - \eta^\prime).
\end{equation}
In \cref{eqn:inverse graviton propagator,eqn:identity} we have omitted $\delta$-functions for the comoving three-momenta. The trace in the flow equation becomes
\begin{equation}
    \tr
    \to \int \dif \eta \dif \eta^\prime \, \delta(\eta - \eta^\prime) \int_{\vec{q}} \Tr.
\end{equation}

The graviton contribution to the flow of the effective potential reads
\begin{equation}\label{eqn:flow due to gravitons}
    \partial_t (a^4 U)
    = \partial_t (A^4 \hat{V})\\
    = -i \int_{\vec{q}} \partial_t R_k(\vec{q},\partial_\eta) \, G(\vec{q},\eta,\eta^\prime)\bigr|_{\eta^\prime=\eta}.
\end{equation}
In \cref{eqn:flow due to gravitons} the local form in $\eta$ reflects the general property in position space for $\Gamma = \int_x L(x)$, where
\begin{equation}
    \partial_t \Gamma
    = \int_x \partial_t \, L(x)
    = \frac{1}{2} \Tr \int_x \int_y \partial_t R_k(x,y) \, G(y,x)
\end{equation}
is solved by the local evolution equation
\begin{equation}
    \partial_t L(x)
    = \frac{1}{2} \Tr \int_y \partial_t R_k(x,y) \, G(y,x).
\end{equation}

We emphasize that both sides in \cref{eqn:flow due to gravitons} involve only frame-invariant quantities \cite{wetterich2016primordial}. In particular, the cutoff function involves the combination
\begin{equation}
    A^2 \hat{k}^2
    = a^4 M^2 k^2,
    \qquad
    \hat{k}
    = a k,
\end{equation}
i.e. $R_k(q)$ depends on $\hat{k}^2$ and $\hat{D}$. Concerning non-linear field transformations the computation of the flow equation is done in a specific frame or choice of fields, namely the one for which the infrared cutoff term is quadratic in the fluctuations. This is the analogue to the selection of a frame in a loop computation by the implicit assumption that no non-trivial Jacobian is present in the functional measure. For our purpose it is important that the cutoff is quadratic in the physical graviton fluctuations, e.g. those that couple to sources reflecting conserved energy-momentum tensors (``linear split''). Once the flow equation has been derived, arbitrary non-linear field transformations can be performed. With respect to conformal field transformations of the metric (Weyl scalings) this is reflected in the frame invariance of the flow equation, e.g. \cref{eqn:inverse graviton propagator,eqn:frame-invariant,eqn:flow due to gravitons}. The solution of the flow equation can be done in an arbitrary frame.

For a flow in Minkowski space with $\tensor{\bar{g}}{_\mu_\nu} = a \tensor{\eta}{_\mu_\nu}$ and constant $a$, one replaces in the previous computation $U \to a^4 U$, $k^2 \to a^2 k^2$ such that $J(q)$ in \cref{eqn:minkowski graviton contribution} involves now $\hat{k}$ instead of $k$. The flow equation for
\begin{equation}
    v
    = \frac{2 U}{M^2 k^2}
    = \frac{2 A^2 \hat{V}}{\hat{k}^2}
\end{equation}
is independent of $a$. More generally, it is invariant under arbitrary $k$-independent Weyl scalings of the metric since the frame-invariant combinations remain unchanged. In other words, the flow equation \labelcref{eqn:flow due to gravitons} is the same for all metrics $\tensor{\bar{g}}{_\mu_\nu}$ that can be transformed into each other by field-dependent but $k$-independent Weyl scalings or conformal transformations of $\tensor{\bar{g}}{_\mu_\nu}$.

For the flow in a curved background geometry one can approximate $\hat{D} = \partial_\eta^2 + \vec{q}^2$ and $\Delta_\gamma = -2 A^4 \hat{V}$ as long as $\hat{k}^2 \gg \hat{\mathcal{H}}^2,\, \partial_\eta \hat{\mathcal{H}},\, \hat{\mathcal{H}} \partial_\eta,\, \hat{K}(\partial_\eta \bar{\chi})^2$. This approximation corresponds to the flow in flat Minkowski space. Indeed, the propagator equation \labelcref{eqn:inverse} for $G$,
\begin{equation}
    \frac{i A^2}{4} \bigl[\hat{D} + R_k(\hat{D}) - 2 A^2 \hat{V}\bigr] G_\text{grav}(\eta,\eta^\prime)
    = \delta(\eta - \eta^\prime),
\end{equation}
can be solved for $\eta$-independent $A$ in Fourier space,
\begin{equation}
    G_\text{grav}(\eta,\eta^\prime)
    = \int_\omega e^{-i \omega (\eta - \eta^\prime)} G_\text{grav}(\omega),
\end{equation}
with $(q^2 = \vec{q}^2 - \omega^2 = a^2 q^\mu q_\mu$)
\begin{equation}
    \frac{i A^2}{4} \bigl(q^2 + R_k(q^2) - 2 A^2 \hat{V}\bigr) G_\text{grav}(\omega)
    = 1.
\end{equation}
Inserting
\begin{equation}
    G_\text{grav}(\eta,\eta^\prime)
    = -\frac{4 i}{A^2} \int_\omega e^{-i \omega (\eta - \eta^\prime)} \bigl(q^2 + R_k(q^2) - 2 A^2 \hat{V}\bigr)^{-1}
\end{equation}
into \cref{eqn:flow due to gravitons}, one recovers the previously discussed flow in Minkowski space. As expected, the integration of modes with large momenta is not influenced by the background geometry if all characteristic length scales of the geometry are much larger than the inverse momentum of the fluctuations.

As $\hat{k}^2$ is lowered and reaches $\hat{\mathcal{H}}^2$, the details of the background geometry influence the flow. A given background geometry could eventually stop the flow once $\hat{k}^2 \lesssim \hat{\mathcal{H}}^2$. This can happen when $\Delta_\gamma$ in \cref{eqn:inverse graviton propagator} vanishes or becomes positive, such that the potential instability is no longer present. Without an instability the graviton contributions are suppressed by $k^2/M^2$ and therefore negligible.

As an illustration, we consider a model with gravity and a scalar field $\chi$. We choose a background metric $\tensor*{\bar{g}}{_\mu_\nu^c}$ and background field $\bar{\chi}_c(\eta)$ that solve the field equations which are derived by variation of the effective action at a given $\bar{k}$. For definiteness we assume for $\Gamma_{\bar{k}}$ the form \labelcref{eqn:vg effective action}. The two independent field equations read \cite{wetterich2016primordial}
\begin{equation}\label{eqn:ife1}
    2 \hat{\mathcal{H}}^2 + \partial_\eta \hat{\mathcal{H}}
    = A^2 \hat{V},
\end{equation}
and
\begin{equation}\label{eqn:ife2}
    \hat{\mathcal{H}}^2 - \partial_\eta \hat{\mathcal{H}}
    = \frac{\hat{K}}{2} (\partial_\eta \bar{\chi})^2.
\end{equation}
For the solution of the field equations \labelcref{eqn:ife1,eqn:ife2} the term $\Delta_\gamma$ in \cref{eqn:delta} vanishes, corresponding to a massless on-shell graviton propagator. In such a geometry, the potential instability for $V > 0$ is cancelled by the geometric terms. The strong IR flow induced by the graviton is no longer present.

The absence of a mass-like term in the ``on-shell'' propagator of the graviton is well known in general relativity and cosmology and has been discussed for the renormalization flow in refs.~\cite{falls2014asymptotic,biemans2017renormalization}. It is at the origin of scepticism about the relevance of the ``avoidance of instabilities'' for observable cosmology. We emphasize that the effective ``on-shell stop'' of the graviton-induced flow occurs only for a particular ``on-shell'' configuration of the scalar field, namely for $\chi(\eta) = \bar{\chi}_c(\eta)$ obeying the field equations. The flow of the effective potential and, more generally the effective action, is an off-shell issue. One evaluates $\Gamma_k$ for arbitrary field configurations, and only a very small subset can obey the field equations derived from $\Gamma_k$. This has important consequences, as can be seen by simple examples:
\begin{enumerate}[label=\roman*.]
    \item We have already found that for $\hat{k}^2 \gg \hat{\mathcal{H}}^2$ the flow is well approximated by the flow in flat space and therefore exhibits the strong graviton-induced renormalization effects.
    \item Consider a potential $U(\chi)$ increasing sufficiently fast with $|\chi|$. For the range $\chi^2 \gg \bar{\chi}_c^2(\eta)$, the potential term will dominate the geometric terms in $\Delta_\gamma$, inducing again a fast flow of $U$ in this region. In variable gravity the asymptotic behavior $U(\chi) \sim \chi^2$ for $\chi \to \infty$ cannot be changed by geometric effects.
    \item For an $\eta$-independent field $\chi$ (instead of $\bar{\chi}_c(\eta)$), as appropriate for the flow of $U$, the term $\sim \hat{K}$ in $\Delta_\gamma$ is missing. (This holds even if $R_k$ employs $\bar{\chi}_c(\eta)$.) Thus $\Delta_\gamma$ remains negative, creating again an instability barrier not to be crossed.
\end{enumerate}

Furthermore, we recall that interesting cosmological solutions are often not simple solutions of field equations for models of gravity coupled to a scalar field. For the matter-dominated universe in the Einstein frame (constant $M$), $a \sim \eta^2 \sim t^{2/3}$, one has $\mathcal{H} = 1/ 2 \eta$, $\mathcal{H}^2 + 2 \partial_\eta \mathcal{H} = 0$ such that the term $\Delta_\gamma = -2 A^4 \hat{V}$ only involves the potential. For the radiation-dominated universe, $a \sim \eta$, $\mathcal{H} \sim 1/\eta$, $\partial_\eta \mathcal{H} = -\mathcal{H}^2$ one infers that $\Delta_\gamma = -2 A^4 \hat{V} - 2 A^2 \mathcal{H}^2$ has a negative geometric contribution, adding to the instability.

In summary, the choice of a realistic cosmological background will typically not modify qualitatively the strong IR effects related to the ``avoidance of instability'' for most parts of the flow. Geometric effects play a role, however, as the flow ends effectively due to the presence of a ``physical IR cutoff'' arising from the geometry of the cosmological solution. The exact flow equation does not depend on the choice of the cutoff. Details of the flow in flat space will, however, depend on the selection of the cutoff. This dependence should be compensated by the details of the stop of the flow due to a physical cutoff, which also depend on the choice of the cutoff. In many circumstances we expect that the role of the geometric effects for a curved background is related to the effective stop of the flow.

\section{Conclusions}

We conclude that the strong infrared renormalization effects induced by the graviton fluctuations are generic for a positive effective potential $U$. They lead to a solution of the cosmological constant problem. The barrier preventing unstable behavior acts by erasing any microscopic value of the cosmological constant, replacing it by a universal scale-dependent infrared value.

Let us concentrate on the action of this ``graviton barrier'' in variable gravity where $M^2(\chi)$ grows for large $\chi^2 \gg k^2$ proportional to $\chi^2$. The flow equation will always induce a dependence of $M^2$ on $\chi$. A scale-invariant coupling $\sim \xi \chi^2 \tilde{R}$ dominates the behavior of $M^2$ for large $\chi$, as compared to any constant contribution. This is consistent with an IR fixed point reached for $k^2/\chi^2 \to 0$, i.e. for $\chi \to \infty$ at fixed $k$. By multiplicative scaling of $\chi$ we can set $\xi = 1$, such that $M^2(\chi) = \chi^2$ becomes indeed a valid approximation. Since the $\chi$-dependence of $M$ is unavoidable, one should derive the flow equations within the framework of variable gravity. Only once the flow equations are derived within variable gravity, they can subsequently be translated to the Einstein frame. As we have argued, this translated flow will not be identical to the flow that one obtains by a direct computation with fixed $M^2$ as performed as a ``warm-up example'' at the beginning of this note. Only qualitative features may be expected to be similar.

Assume that the cosmon potential $U(\chi)$ increases for increasing $\chi^2$, with $U(\chi) > 0$. The strong graviton-induced flow limits the asymptotic increase of $U(\chi \to \infty)$ to $U \sim k^2 \chi^2$ - in other words, the graviton barrier blocks any faster increase. As a consequence, the dimensionless ratio $\hat{V}(\chi) = U(\chi)/M^4(\chi) = U(\chi)/\chi^4$ decreases for increasing $\chi^2$, $\hat{V} \sim k^2/\chi^2$. For cosmological solutions with $\chi^2 \to \infty$ for the infinite future, the observable cosmological constant vanishes asymptotically \cite{wetterich1988cosmology,wetterich2015inflation}. If $U(\chi \to \infty)$ increases more slowly than $\chi^2$, or goes to a constant, the decrease of $\hat{V}$ with $\chi$ is even faster. An increase of $|U(\chi \to \infty)|$ with $\chi^2$ towards negative values is not compatible with a potential bounded from below, as required for a consistent quantum field theory. We conclude that the graviton barrier implies an asymptotic decrease of the dimensionless cosmological ``constant'' $\hat{V}(\chi \to \infty)$ to zero. An asymptotically vanishing cosmological constant obtains in more general settings as well. It is sufficient that $\hat{V}(\chi) = U(\chi)/M^4(\chi)$ vanishes for $\chi \to \infty$.

Once the flow equation for $A^4 \hat{V}$ is derived within variable gravity, we can transform this description into the Einstein frame with $\chi$-independent $M^2$. This is achieved by a $\chi$-dependent Weyl scaling of the metric. As long as the Weyl scaling is $k$-independent, the exact flow equation is the same for all frames. We have formulated the flow in terms of frame-invariant quantities such as $\hat{V}$. In the Einstein frame, $U = \hat{V} M^4$ decreases $\sim \chi^{-2}$ and goes precisely to zero (not a non-zero constant!) for $\chi \to \infty$. For a rescaled scalar field $\varphi$ with canonical kinetic term, the decrease of $U$ is approximately exponential, $U \sim M^4 \exp(-\alpha \varphi/M)$. Such potentials give rise to dynamical dark energy or quintessence \cite{wetterich1988cosmology}, typically inducing ``scaling'' or ``tracking'' solutions for cosmology \cite{wetterich1988cosmology,ratra1988cosmological}.

We infer that dynamical dark energy is a rather natural consequence of the graviton barrier. Since $\hat{V}$ approaches zero for $\chi \to \infty$ with $\hat{V} > 0$ for finite $\chi$, the only possibility to avoid dynamical dark energy is a minimum of $\hat{V}$ for a finite value $\chi_0$. This requires a maximum for some finite $\chi_\text{max} > \chi_0$ as well. While this remains a possibility, the case of monotonic $\hat{V}$ appears to be simpler. (We observe that in the case of a minimum, the value $V = U(\chi_0)$ would also be subject to strong IR renormalization effects if $V > 0$.) Our computation of the graviton-induced renormalization effects are in support of the ideas underlying the first proposal of quintessence \cite{wetterich1988cosmology}, of crossover variable gravity \cite{wetterich2015inflation} and of the prediction of the Higgs mass within asymptotic safety \cite{shaposhnikov2010asymptotic}.

We have only briefly discussed curved background geometries. Under certain circumstances, they may stop the IR flow for low $k$ if metric and scalar fields are close to solutions of the field equations or if geometry provides an effective infrared cutoff in some other way. A quantitative understanding of this effect will require some technical effort in order to derive and solve flow equations in a time-dependent background. In view of our qualitative findings for the overall flow, we conclude that curved background geometries do not change our main conclusion: strong infrared quantum gravity effects solve the cosmological constant problem.

Interesting open questions emerge: Which are the effects of the graviton barrier for other parts of the effective action for gravity, in particular in the low momentum domain? Are the infrared gravity effects consistent with the numerous precision tests of gravity which confirm the Einstein-Hilbert action? Could there be observable consequences, or implications for black holes? Functional flow equations evaluated on the corresponding geometries should be able to address these questions. This should also shed light on more phenomenological investigations \cite{bonanno2002cosmology,reuter2005big,sola2017towards} of possible consequences of the renormalization flow of the cosmological constant. If these issues show no conflict with observation, our findings solve the fundamental puzzle why our universe has grown large enough in a not too disruptive way, such that complex structures as galaxies, stars and life could emerge. The border of instability gives room for complexity.

\paragraph{Acknowledgement} The author would like to thank A. Eichhorn, S. Floerchinger, H. Gies, J. Pawlowski, R. Percacci, M. Reuter and F. Saueressig for comments and discussion. This work is supported by ERC-advanced grant \href{http://cordis.europa.eu/project/rcn/101262_en.html}{290623} and the DFG Collaborative Research Centre ``\href{http://www.dfg.de/en/research_funding/programmes/list/projectdetails/index.jsp?id=273811115&sort=var_asc&prg=SFB}{SFB 1225} (ISOQUANT)''.

\printbibliography

\end{multicols}

\end{document}